# Signal Diffusion Mapping: Optimal Forecasting with Time Varying Lags.

Paul Gaskell – University of Southampton, WAIS
Frank McGroarty – University of Southampton, Finance and Banking Group
Thanassis Tiropanis – University of Southampton, WAIS

**Abstract**

We introduce a new methodology for forecasting which we call Signal Diffusion Mapping. Our approach accommodates features of real world financial data which have been ignored historically in existing forecasting methodologies. Our method builds upon well-established and accepted methods from other areas of statistical analysis. We develop and adapt those models for use in forecasting. We also present tests of our model on data in which we demonstrate the efficacy of our approach.

**Introduction**

Conventional time series methodologies for financial forecasting are limited in their ability to handle lags. They can only accommodate lags fixed integer length, e.g. a one-period lag, a two-period lag, etc.. We argue that the relationship between real time series can often have a richer, more dynamic time structure than has been implicitly assumed in the traditional methodological approaches. We introduce a new forecasting methodology, Signal Diffusion Mapping (SDM), which extracts the maximum possible information from the modelled relationship and produces the best possible forecast at each point in time, in circumstances where the lag-length is time-varying.

The notion of time-varying lag-lengths may strike some researchers as odd. Indeed, we suspect that the fact that lags have been almost invariably modelled as fixed-integer-lags in time-series financial modelling, has probably conditioned most financial forecasters to believe that financial data actually behaves in the manner that their models seek to measure. However, we contend that that such a conclusion would be a classic case of "if all you have is a hammer, everything looks like a nail".

Real world temporal relationships are less well ordered than we might like to think. The history of financial markets is replete with colourful examples of convicted insider traders who exploited access to privileged information for pecuniary gain.  This can include information about companies' earnings announcements, future takeover targets, price-sensitive macroeconomic news, etc.. An insider trader trading prior to a news event will reverse the time-order that financial theory conventionally assumes, i.e. information first, price-reaction second. On the other hand, nobody would accept the assertion that insider-trading occurs prior to each and every information event. In other words, a reasonable conclusion would be that occasional cheating happens and this distorts the information-price time-lag. If a fixed-integer-lag model were imposed on such a scenario, it would results in the measured correspondence between information and price being lower than if the modeller had a means of adjusting for the time-distortion.

Insider trading is not the only source of lag-length distortion. Cheung et al (2004) survey the opinions of foreign exchange dealers and conclude that "news", "speculative forces" and "bandwagon effects" are the main drivers of price within a day and that fundamentals, which mainstream finance theory assumes to be the key drivers of price, are perceived as only relevant in determining returns over the long term (i.e. over 6 months). These speculative forces and bandwagon effects are perhaps best captured in Soros' (2008) reflexivity theory which posits a bi-directional symbiotic feedback relationship between information and price. Importantly, Soros (2008) argues that some reflexivity relationships are sustained for extended periods, while others can fizzle out after a short time. Trying to extract the relationship between such variables with a fixed-length-lag model is like trying to eat soup with a fork – quite simply the wrong tool for the job.

The key contribution of this paper is the introduction of a new model (SDM) to the forecasters' arsenal which can handle time-varying lag-lengths of the type we describe above, retrieving the maximum possible information about the relationship between two time series. Our method builds upon well-established, published methods from the data analysis and econophysics literatures, which were developed to measure optimal relationships in related series of fixed lengths, e.g. Keogh and Pazzani (1999), Sornette and Zhou (2005), Zhou and Sornette (2007). However, in their original form none of those methods were appropriate for forecasting. We adapt those methods to the problem of forecasting in a manner analogous to a Kalman Filter or Recursive Bayes Estimation with time-varying-lags, which we name Signal Diffusion Mapping. We test SDM with synthetic data and we demonstrate that it is able to recover the true optimal relationship between series where we have deliberately distorted the temporal relationship by shortening/lengthening the lag at different points in time.

Please note that SDM does not rule out the possibility that the true relationship between two time series could actually be a fixed length lag. An important feature of our proposed methodology is that it will identify the optimal underlying relationship between the two series whether the lag structure is fixed of varying in time. In other words, if the best possible relationship is obtained by lagging one of the series by two time periods, then that is what the model will find. On the other hand, if an even better link between the two series could be shown by allowing series 1 to lead series 2 some of the time and to lag series 2 at other times, our method will identify this as the best model.

The remainder of our paper is structured as follows: Section 1 introduces some concepts and notations from the relevant literatures we will use in defining our approach. Section 2 will describe the SDM algorithm in full. In section 3, we present results from testing the SDM algorithm on simulated data.

Finally, section 4 presents our conclusions.

## 1. Background

In the introduction we highlight the work of Soros (2008) and Cheung et al (2004), as examples of work which highlights the temporal complexity of the information-price relationship. In the case of Soros, this complexity is captured as a constantly shifting feedback mechanism, capable of rapid changes – for example periods of constant growth lasting years can shift to huge market crashes in the space of less than a week. In the case of Cheung et al the emphasis is on scale, where short term information effects (intra-day) are augmented with long term effects of 6 months or more.

As described, this implies the information-price relationship has a complex causality structure. Statistical causality in the financial literature is mostly considered in terms of Granger causality (Granger 1969), so that where the time-sequencing of a relationship is unclear, Granger causality tests are employed as the main tool in determining the nature of the statistical causality.

In section 1.1 we are going to introduce the Granger causality model and cases where it has been applied in the study of information-price relationship. We will also discuss some other approaches to determining statistical causality. We are then going to discuss the limitations of these models in terms of the dynamism and scaling of the relationship we wish to measure.

In section 1.2 we will discuss how the Bayesian framework is a good fit for problems of this type. We will briefly highlight other work where Bayes Estimators have been used for model fitting in a financial context and draw the link to other state space techniques used in time-series analysis.

### 1.1. *Time-Sequencing in Financial Forecasting*

To give the issue of variable lag-lengths a mathematical framing, the standard relationship that is considered in the Granger causality approach is that some autoregressive series of the form.

$$x_t = A x_{t-1} + N(0,1)$$

Where A is the autoregressive parameter, |A| < 1 so the process is stationary and N(0,1) indicates a random variable with mean 0 and standard deviation of 1. We then consider another variable related to y as;

$$y_t = \beta_1 x_{t-1} + \beta_2 x_{t-2} ... + \beta_{t-\tau} x_{t-\tau} + u_t \quad \text{where} \quad u_t = N(0, \sigma_u) \text{, (1)}$$

is the noise term obfuscating the relationship.

Granger causality in the bi-variate case is simply a regression model specified on the lagged values of **x**, throughout this paper bold-face type will denote a vector in the usual manner, where β are the coefficients values or parameters of the model, in some cases we might also consider lags of **y** but here we omit them for brevity. The parameters of the model are then estimated with the Ordinary Least Squares algorithm. If any of the β coefficients are statistically significant, researchers can then state that there is evidence to support the hypothesis that **x** Granger causes **y**.

This is by far the most common form of analysis of the lagged dependence between time-series, example studies of information price relationships using this methodology include; Bollen (2011) and Sprenger (2013), both studying the effect of public sentiment on stock prices, Hong et al (2009) modelling risk spillover between international financial markets and Hiemstra and Jones (1994) studying the causality between stock prices and volume using the Granger causality test specified in (1) and a non-linear variant.

Even in cases where more complex methodologies are applied, usually the treatment of the temporal relationship between the variables reduces to some variant of Granger causality. An example of this is Sharkasi et al (2005), who use wavelet decomposition to study spillover effects from international stock indices, then use linear regression to fit models on the decomposed wavelets.

One of the attractions of the Granger causal approach is that it gives a clear pathway to constructing a forecasting model. Based on the results of (1) it is possible to identify lags with significant p-values, then construct a second regression using these lags. This second model can then be used to determine estimates of future values of **y** based on the coefficient values in the second regression model.

The situation Soros (2008) and Cheung et al (2004) describe, however, is one of significant variation in the temporal relationship between variables, both in terms of the speed by which the relationship can change but also the fact that the relationship can exist at varied scales. In Soros' description of reflexivity he states how financial markets can switch from bull to bear periods very quickly, these switching moments are characterised by changes in the information-price relationship that happen over the course of days - suggesting any model would have to adjust almost immediately to the change in circumstance. Cheung et al report how professional traders consider information effects at different scales, from intra-day effects to long run effects, >6 months.

Now consider trying to capture this with the Granger causal approach. Firstly tackling scale; Cheung et al's conclusions indicate we would need to fit a model with a large number of parameters to capture each different scale considered. Consider daily intervals for example, this would lead to a model with ~180 parameters, leading to very low statistical power for the test due to the 'curse of dimensionality'. Secondly, assuming the parameters of (1) are time-varying, we could employ a version of (1) on some rolling window. In order to achieve reasonable results, however, requires specifying a reasonable sized window – yet the theory suggests that change in the relationship will occur almost instantly.

1.2. *Parameter Estimation using Bayesian Inference*

To be specific; the issue we have is that given the Granger causal model we believe the model specification would have too many parameters - and that these parameters would change too quickly, for us to be able to fit the model we need to test the available financial theory with the Ordinary Least Squares (OLS) algorithm. We are not arguing that the basic structure of the Granger causal model is deficient in it's ability to characterise the information-price relationship *per se*, just that we cannot fit the model in the way we would like.

Recently, there have been a number of papers in the finance and econometrics literatures using various types of Recursive Bayes Estimator (RBE) to estimate model parameters. Examples of this work include, Carvalho and Lopes (2007) who present an RBE for dynamically parametrised stochastic volatility models and Carvalho et al (2009) who fit the parameters of a dynamic, conditionally linear model (see Arulampalam *et al* 2002 and Lopes and Tsay 2011 for reviews of the field). As yet, however, no work exists attempting to fit causality models in a similar fashion.

From a Bayesian perspective the parameter values for the lags in (1) can be considered as a state space – where the parameters are represented as probabilities that a given lag could be causally influencing the value of $y_t$. Simplistically this means that we can update the parameters based on their relative probabilities as they transition from one time-period to another, rather than on their goodness of fit over a large number of previous observations. We will show how this can allow the lag-structure to vary much more dynamically than it could do if fitted with OLS and circumvents issues of dimensionality.

This state space representation also links to other areas of the time-series analysis literature where variable lag paths have been considered. The Dynamic Time Warping (DTW) (see for example Keogh and Pazzani 1999, Senin 2008, Warren Liao 2005, Sakurai et al 2005) and Optimal Thermal Causal Path (OTCP) (Sorentte and Zhou 2005, Zhou and Sornette 2007) literatures both study historical lagged relationships using similar state space techniques. Much of section 3.2 concerns integrating these ideas into the RBE framework - so that we can forecast with variable lag-structures, rather than view them historically.

The key contribution we make in this paper, is to show it is possible to dynamically fit the model (1) using a RBE, we will begin by fitting the simple linear model and expand to more exotic cases in later sections. The specific RBE algorithm we present to complete this task is the Signal Diffusion Mapping (SDM) algorithm. The rationale for the name is because we will map the diffusion gradient of information flowing between the two series over the state space of possible lags – this results in us being able to plot the time evolution of this relationship on a heatmap – i.e. the map of the flow of the signal between the series.

In section 2 we will further define the time-sequencing problem in Bayesian terms and introduce the notational conventions we will adhere to throughout this paper.

## 2. A Bayesian View of the Time-Sequencing Problem

In this section we are going to interpret the Granger causal model in terms of the common General Dynamical Model (GDM), this a more general form of the Normal Linear Dynamical Model which forms the basis of the Kalman Filter (Lopes and Tsay 2011). We will then outline the general method of solving the GDM equations recursively using Bayes theorem. Finally we will describe the criteria by which an estimator can be seen as optimal which will leave a clear pathway to introducing SDM as the optimal solution to these equations.

The Bayesian approach to statistical decision making is based around the definition of relative beliefs about a set of different events. These beliefs are represented as a state space of probabilities associated with each possible event - given (1) our beliefs are about a series of discrete causal relationships between **x** and **y**. We are going to hold this set of beliefs in a probability vector containing an entry for each of the lagged values under consideration - we denote these options as an Ns length probability vector **w**$_t$ where Ns is the number of considered lags (read 'number of states') at t, i.e.;

$$w_t : [w_t^1 ... w_t^{Ns}] \qquad \text{and} \qquad 1 = \sum_{i=1}^{i=Ns} w_t^i .$$

The convention we adhere to throughout this paper is that subscripts will denote a vectors position in time and superscripts denote the relative position of a value in the vector, so $w_t^i$, would be read the i'th position on vector **w** at time t. In terms of (1), these probability weights are conceptually similar to the parameter values β – we differentiate them in the notation to save confusion as they are probability weights rather than regression coefficients and will further define this difference in section 3.2.

The second aspect of the Bayesian approach is to then define a measure of how well these beliefs map on to some measurements of the processes under study. Let **d**$_t$ be an Ns length measurement vector $d_t : [d_t^1 ... d_t^{Ns}]$ containing some measure of the **x**, **y** relationship corresponding to each of the values of **w**$_t$. For now the reader simply has to understand this as a comparative measure of the relationship, we will make this definition concrete in the following section.

We note that it is typical in the RBE literature to use x for the state vector and y for the measurement vector. We have differed from this notation because in the econometrics literature x and y are typically the time-series under study. As we believe SDM is mainly aimed at the financial forecasting community we have sided with the econometric notational convention.

The GDM is then given by two equations – the first, usually referred to as the *system model,* governs the way in which our beliefs about the system propagate forward through time, this is defined in probabilistic terms as;

$$w_t \propto p(w_t|w_{t-1}) \quad (2)$$

where the operator should be read as 'varies in proportion to' - so the equation states that the probability densities of **w**$_t$ vary proportionally to a probability mass function applied to the densities at t-1.

The second, usually called the *measurement model*, governs the way we are going to interpret the measurement vector in terms of the state probabilities;

$$d_t \propto p(d_t|w_t) \quad (3)$$

which should be read that the measurements vary proportionally to the likelihood of the measurement given the state vector probabilities.

If the system model can be characterised as a Markov chain i.e. the values of **w** at t+1 depend only on the values at t, then we can define a Bayesian prediction model for the updated values of **w**$_t$ as;

$$w^i_{t|t-1} = \sum_{j=1}^{j=N} p(w^i_t|w^j_{t-1}) w^j_{t-1} \quad (4)$$

Where $w^i_{t|t-1}$ denotes the estimate of the probability density of the i'th position on the probability vector **w** given the prior densities and the system model. All that (4) really states is that, if we have a system model holding the transition probability of the i'th lag holding useful information given the preceding vector of lags, then we can sum over these probabilities to estimate the next t's value for the lag. It is common in the literature to see (4) written as an integral, the reason for the summation in our case is that the state space over the lags is discrete, rather than continuous.

Given an estimation of the weightings of **w** based on the known densities at t-1 and the system model. We then receive a set of measurements **d**$_t$ that we use to update this forward projection of the densities based on the observed evidence. Given the measurement model (3) we can write this due to Bayes' theorem as;

$$w^i_t = \frac{p(d^i_t|w_t) w^i_{t|t-1}}{\sum_{j=1}^{j=Ns} p(d^j_t|w_t) w^j_{t|t-1}} \quad (5)$$

Which gives the likelihood of the measurement given the data, multiplied by the prior probability if the i'th lag normalised over all Ns possible lags.

This yields the basic prediction and update structure of a *Grid Based Filter*, which is a type of RBE defined on a discrete set of possible states – in this case lags, and forms the basis of a number of RBE algorithms, two of the best known being the Bootstrap Filter (Gordon et al 1993) and the Auxiliary Particle Filter (Pitt and Shephard 1999). The attraction of this formulation is that, providing we can valid functional forms for (2) and (3), the filter will evolve to the optimal calculation of the probability densities of **w** over repeated iterations of (4) and (5) (Arulampalam *et al* 2002, Lopes and Tsay 2011).

Optimality in this sense means the probability densities which minimise the measurement error. So we need a measurement model which is consistent with reasonable assumptions about the **x**, **y** relationship. We also need to specify a probability density function for the values of these measurements, since (5) requires that we calculate the likelihoods for these observations given our beliefs about lag probabilities. Finally, we require a probability mass function for (3) which captures the time-evolution of the lag structure in a theoretically justifiable way.

If these criteria are met, we can claim the estimator is optimal under the specified assumptions. The trade-off in defining the estimator is then how to posit relaxed enough assumptions about the forms of (2) and (3) to make the estimator generally applicable to a range of forecasting tasks.

What we are going to show in introducing SDM is that we can specify the form of (2) using very weak assumptions. These assumptions are well supported by other areas of the literature. This effectively removes time-varying lags as an issue and allows us to plug in any relevant measurement model available in the econometrics literature. This makes the SDM algorithm useful in a large number of practical applications, since the researcher can still utilise all of the current modelling approaches for bivariate series and plug in the SDM estimator as the time-sequencing test supplementing the specified model.

## 3. Signal Diffusion Mapping

In the following subsections we will being by describing the system model we are going to use for (2). This is the key operation in the SDM algorithm and remains invariant despite the specification of the measurement model. We will then define a measurement model to fit the simple Granger causal model (1). After illustrating this simple case we will then go on to describe more exotic variants of the SDM algorithm i.e. bi-directional causality structures and positive-negative switching causality structures. Finally, we will give an algorithmic implementation of the SDM algorithm and discuss implementation issues and computational complexity.

3.1. *System Model*

From a forecasting perspective, clearly we would like to be able to define a model where the densities over the lags do not change over time, in this case we would be able to estimate *when* the relationship was likely to occur perfectly. Another way of stating this would be that the error in the forward projection of our beliefs in the systems state is 0 and $w_t = w_{t-1}$.

If there is variation in our beliefs over time then we need to posit an equation describing this variation. Here, there are two important quantities. Firstly; the structure this variation takes – i.e. how this variation deforms the densities of **w**. Secondly; the magnitude of the variation – i.e. how much deformation in the density vector we expect. After defining this function we then need to find the parametrisation of the function which minimises the total error for the system model over time.

In terms of the structure of the variation, there is already a large body of literature dealing with variations in historical lag structures. We are going to follow these literatures in defining the structure of the temporal variation similarly – just as a probabilistic forward projection of the state vector probabilities, rather than a historical representation.

Both the Dynamic Time Warping (DTW) (see for example Keogh and Pazzani 1999, Senin 2008, Warren Liao 2005, Sakurai *et al* 2005) and Optimal Thermal Causal Path (OTCP) (Sorentte and Zhou 2005, Zhou and Sornette 2007) literatures use state space methods to study historical lead-lag relationship between time-series. The basic premise of either algorithm is to define a matrix of all of the pairwise relationships between the variables on a matrix containing a measure of the relationship in each square. Then traverse the matrix from a fixed start to fixed end point in such a way as to reveal the optimal – or lowest cost relationship between the variables based on the measure.

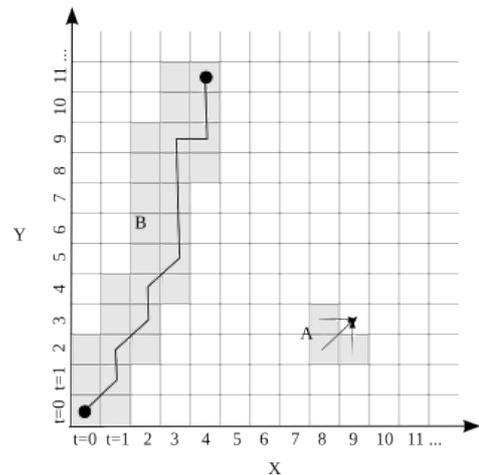

In order to do this, both algorithms make the assumption that the structure of the lagged relationship can vary only slowly in time. Slow varying in this sense means that a lag path can vary by only 1 unit time-period for every unit increase in t.

Taking DTW as an example (although both algorithms are derived from the same premise), given two time-series $c:[c_1 ... c_k]$ and $q:[q_1 ... q_k]$, we can construct a k*k matrix of the pairwise distances, i.e. $|c_i - q_j|$ between the two series for each lag length. The DTW algorithm then seeks the lowest distance, continuous 1 to 1 mapping between the two series from a fixed start point to a fixed end point.

*Figure 1: Slow Varying Lag Paths; (A) describes the three squares in the matrix via which we can reach square (9, 3) due to the recursive relation used in the DTW and OTCP algorithms. (B) then shows an example of a path defined as a continuous mapping between points (0,0) and (4,11).*

This task is completed by understanding that the lowest distance path to a point on the matrix must be the summation over the lowest distance pathways up to this point – which justifies the recursive relation;

$$\varepsilon_{i,j} = |c_i - q_j| + min[\varepsilon_{i-1,j}, \varepsilon_{i-1,j-1}, \varepsilon_{i,j-1}] \quad (6)$$

where $\varepsilon_{i,j}$ is interpreted as the relative cost of the pathway – so that the minimum distance pathway is the lowest cost pathway up to a given point on the matrix.

Defined in this way the lag path has some desirable properties we would expect in a realistic lagged relationship;

1) A relationship is defined for each t – there are no large jumps where no relationship is defined for a set of time-periods.
2) The relationship is a continuous 1 to 1 mapping – if this was not the case there could be overhangs or cliffs in the lag path implying the causal relationship ran from the future into the past.

(adapted from Sornette and Zhou 2005)

The way we are going to implement these ideas in the SDM estimator is to make the same assumption – that the time evolution of the lag structure is relatively slow. Given that the ideal forecasting scenario is that our beliefs about the lag probabilities are accurate this means characterising any deviation, or error, in these beliefs as a relatively slow varying function of the state vector.

In our notation, (6) implies allowing a given value of $\mathbf{w_t}$ to be influenced by one of the following 3 values;

$$(w_{t-1}^{i-1}, w_{t-1}^{i}, w_{t}^{i+1}) \quad (7)$$

Assume that we knew the amount of variation in the lag structure with certainty, so we could fix a parameter θ, which captured the total magnitude of the variation. Given that we are expecting $w_t = w_{t-1}$ in the 0 error case, (7) tells us the structure of the variation so we could substitute into (2) and write;

$$w_t \propto w_{t-1} + error \qquad \text{so} \qquad w_t^i \propto w_{t-1}^i + \frac{\theta}{3}(w_{t-1}^{i-1} + w_{t-1}^i + w_t^{i+1}) \quad (8)$$

Dividing the magnitude of the error by 3 is due to the fact that we are distributing this quantity evenly over the lags available due to the slow varying constraint.

Clearly in most interesting cases we are not going to be able to assume a constant amount of variation, so a better characterisation would be as a random sequence of perturbations in the lag structure $v:[v_1 ... v_t]$. We could define a mass or density function for v but as we will show this is not necessary and $\mathbf{v}$ can be any arbitrary sequence. We would then like, at each t, to choose a parameter value for θ which minimises the global error over repeated iterations i.e. minimises the amount of variation in our beliefs about the lag structure.

To do this we can define some properties of v due to the properties of the probability vectors. Firstly, we note that $0 \leq |w_t - w_{t-1}| \leq 2$, so assuming no error we can also state $|w_t| - |w_{t-1}| = 0$. It follows then that the magnitude of the error process at t is the distance between the vectors $v_t = |w_t - w_{t-1}|$. Further, we can also infer that $v_t$ always has a expected value, since the expectation of a positive random variable is always defined, and that this expectation must lie on the bounded interval [0, 2].

Under these circumstances, the optimal choice of value for θ is the median of the distribution of the errors up to this point – this is due to the fact that the median is the minimiser of the distance function where the expected value of the function is defined - so we can define the optimal choice of value for θ as;

$$\theta_t = \tilde{v}_{1:t-1} \quad (9)$$

Where the tilde denotes the median of the vector of observed errors.

Substituting into (4) then yields the system model for the forward projection of the state vector as;

$$w_{t|t-1}^i = w_{t-1}^i + \frac{\theta_t}{3}(w_{t-1}^{i-1} + w_{t-1}^i + w_t^{i+1}) \quad (10)$$

An important note is that at the first and last positions, i.e. the boundaries, on **w** not all of the 3 positions will be defined since we have no data for either $w_{t-1}^{1-1}$ or $w_t^{N+1}$, here we simply set the probability of these positions to 0. A further point of note is that (8) implies $\sum w_{t|t-1}$ takes values of > 1 where θ > 0, so is no longer a probability vector. At this stage a better interpretation of these estimates is as a set of weights defined by their relative likelihoods. We could normalise this vector to 1 by dividing over the sum of the weights but this is unnecessary as we will compute the actual probabilities using the update step (5).

In terms of the optimality conditions we state at the end of section 2, (10) is always defined for any pair of series irrespective of their spatial relationship. This follows because the information about the spatial relationship between the series is held as probabilities – so for the reasons already specified we can always calculate the median of the distribution and this will always be the optimal estimate for θ$_t$. The assumption we make is that the time-evolution of the causality structure is relatively slow, this assumption is justified by a the large literature on DTW algorithms and the nascent OTCP literature.

The attractiveness of this approach is it allows us to concentrate on modelling the complexity of the distance relationship between the series independently of the temporal relationship – to show this we will describe how to fit (1) using this system model in the next section, then expand to a number of other cases.

3.2. *Measurement Models*

To complete the SDM estimator we need to define a measure of the x, y relationship and posit a functional form for mapping these measures into the same probability space as the system model. This means making some assumptions about the form of the distances between the two series then using these assumptions to calculate the likelihood of the measurement. We than substitute these likelihoods into (5) and the estimator is complete. We will begin in section 3.2.1 by describing how to fit (1) where x and y are assumed to be positively correlated. We will then show, in subsequent sections, how we can relax these assumptions so that we can fit bi-directionally causality structures and models where there are positive-negative regime shifts in the causality structure.

3.2.1. *Simple Linear Causal Models*

If we assume that x and y are measured in the same units, for example they may have been rescaled using their respective means and standard deviations, then if there was 1 lagged value of x causally related to y we could write this as;

$y_t = x_{t-\tau} + u_t$ (11)

If there is more than 1 lagged value of x causally related to y the OLS approach to model fitting is to interpret further lags as independent random variables. So the model expands by adding more random variables into the regression. The difference in approach using SDM is that we are going to maintain the assumption that there is only 1 causal relationship between the variables but that this relationship is distributed over a number of lags. In this way we can think of the probabilities as a weighted average over a partitioned interpretation of x i.e.

$y_t = u_t + \sum_{i=1}^{i=Ns} w_t^i x_{t-i}$ (12)

We are going to use the squared distance between x and y as the measure of the relationship, so that in effect the SDM estimator becomes the least squares estimator for a relationship with time-varying lags. Squaring the terms in (12) then yields.

$y_t^2 = \left(u_t + \sum_{i=1}^{i=Ns} w_t^i x_{t-i}\right)^2$ (13)

Which we can rearrange to;

$$u_t^2 = \sum_{i=1}^{i=Ns} w_t^i (x_{t-i} - y_t)^2 \quad (14)$$

Then setting the squared distance as the measure of the relationship so that $d_t^i = (y_t - x_{t-i})^2$ we can write (14) in the more compact form;

$$u_t^2 = \sum_{i=1}^{i=Ns} w_t^i d_t^i \quad \text{or} \quad u_t^2 = w_t \cdot d_t, \quad (15)$$

where the dot indicates the scalar product of the two vectors. The assumption of the model (1) is that u is a mean 0 random variable with unknown variance. As a result we would expect the squared values of u to be drawn from the Gamma distribution with a shape parameter of 1. The density function for this distribution is given as.

$$f_{U^2}(u^2|\lambda) = \lambda e^{-\lambda u^2} \quad (16)$$

Where λ is the rate parameter with the maximum likelihood estimator given as the mean of the observed values of $u^2$ i.e.

$$\lambda_t = \frac{1}{\overline{u_t^2}} \quad \text{where} \quad \overline{u_t^2} = \frac{1}{t} \sum_{s=1}^{s=t} w_s \cdot d_s . \quad (17)$$

We have included the time-varying subscript for λ as we are going to calculate the parameter of the distribution through repeated samples over time. Given an estimate for λ we can then calculate the likelihood of a given distance causally influencing the values of y using the likelihood function conditional on the value of λ.

$$p(d_t^i|\lambda_t) = \lambda_t e^{-\lambda_t d_t^i} \quad (18)$$

Substituting these likelihoods into the system model equation (5) then yields.

$$w_t^i = \frac{p(d_t^i|\lambda_t) w_{t|t-1}^i}{p(d_t|\lambda_t) \cdot w_{t|t-1}} \quad (19)$$

Note that different to (5) we have used the scalar product notation for the denominator of (19). Combining (5) and (19) will then give the the optimal recursive estimator of (1), under the specified assumptions.

3.2.2. *Bi-directional Causality Structures*

In the introduction we discussed certain occasions where the time-ordering of information-price relationships may be reversed – i.e. when insider trading occurs in the run up to an information event. The estimator we have presented so far only considers causality running from information to price. In this section we will show how it is easy to generalise the SDM estimator to cases of bi-directional causality.

A simple bi-directional analogue to (1) is the system of equations.

$$y_t = \beta_1 x_{t-1} + \beta_2 x_{t-2} \ldots + \beta_{t-\tau} x_{t-\tau} + u_t, \quad u = N(0, \sigma_u)$$
$$x_t = \alpha_1 y_{t-1} + \alpha_2 y_{t-2} \ldots + \alpha_{t-\tau} y_{t-\tau} + \mu_t, \quad \mu = N(0, \sigma_\mu) \quad (20)$$

As there are now probability weightings associated with the lagged values of either variable, to simplify the notation we introduce the probability matrix **W** indexed at t, but containing a column of values for either series. The first column will contain the lagged values of x and the second the lagged values of y so that the entry $w_t^{i,1}$ would indicate the i'th lag of x at time t and $w_t^{i,2}$ indicates the i'th lag of y at time t so that;

$$W_t = \begin{pmatrix} w_t^{1,1} & w_t^{1,2} \\ w_t^{2,1} & w_t^{2,2} \\ ... & ... \\ w_t^{Ns,1} & w_t^{Ns,2} \end{pmatrix}$$ where since W is a probability matrix $\sum W_t = 1$ . (21)

We are going to then define the distance measures associated with each of the lagged values in a matrix with corresponding entries to **W** i.e.

$$D_t = \begin{pmatrix} d_t^{1,1} & d_t^{1,2} \\ ... & ... \\ d_t^{Ns,1} & d_t^{Ns,2} \end{pmatrix}$$ where $d_t^{i,1} = (y_t - x_{t-i})^2$ and $d_t^{i,2} = (x_t - y_{t-i})^2$

Due to (15) we can then write the squared distances for system of equations described in (20) as.

$$u_t^2 = \sum_{i=1}^{i=Ns} W_t^{i,1} D_t^{i,1} \quad \text{and} \quad \mu_t^2 = \sum_{i=1}^{i=Ns} W_t^{i,2} D_t^{i,2},$$

which implies;

$$(u_t + \mu_t)^2 = (\sum_{i=1}^{i=Ns} W_t^{i,1} D_t^{i,1}) + (\sum_{i=1}^{i=Ns} W_t^{i,2} D_t^{i,2}) \quad \text{or} \quad (u_t + \mu_t)^2 = W_t \cdot D_t, \text{ (22)}$$

using the scalar product notation. Since the summation of two Gaussian distributions is another Gaussian distribution, the maximum likelihood estimator for the sum of the squared error terms u and μ is the mean of the globally observed errors - due to (17) we can then write this as.

$$\lambda_t = 1/(\frac{1}{t} \sum_{s=1}^{s=t} W_s \cdot D_s) \text{ (23)}$$

The system model is then applied independently to either column of the matrix and the measurement model is applied to the **D** and **W** matrices using**.**

$$W_t^{i,j} = \frac{p(D_t^{i,j}|\lambda_t) W_{t|t-1}^{i,j}}{p(D_t|\lambda_t) \cdot W_{t|t-1}} \text{ (24)}$$

The estimator is again optimal under the same assumptions as the uni-directional case.

### 3.2.3. *Positive-Negative Regime Shifts*

The formulation of the estimator we introduce in section 3.2.2. shows how we can allow different functional representations of the x, y relationship to compete against each other for a share of the global probability density. Another case we might consider is where there is uni-directional causality running from x to y, but there are regime shifts from positive to negative correlation in the nature of the relationship. To capture this structure we can define a distance matrix containing the positive and negative squared distances so that;

$$D_t = \begin{pmatrix} d_t^{1,1} & d_t^{1,2} \\ ... & ... \\ d_t^{Ns,1} & d_t^{Ns,2} \end{pmatrix}$$ where $d_t^{i,1} = (y_t - x_{t-i})^2$ and $d_t^{i,2} = (y_t + x_{t-i})^2$

The rest of the estimator is constructed in the same way as the bi-directional case described in section 3.2.2. and remains optimal under the same assumptions.

### 3.3 *Algorithmic Implementation*

Up to this point we have assumed that we would include every lag of **y** in the forecast of **x**. In most forecasting scenarios this would lead to a large tail of extremely low probability lags as more time-periods are considered. It also means that the computational complexity of the SDM algorithm would scale exponentially with t. Clearly in most forecasting tasks we only wish to consider a finite number of lagged values of the either series and so we can bound Ns to some reasonably small number, in this case the algorithm scales linearly with t. Consider the following pseudo-code implementation of the algorithm for the simple uni-directional linear causality model described in section 3.2.1.

*Pseudo Code*

$inputs: \mathbf{x}[x_1...x_t], \mathbf{y}[y_1...y_t], \mathbf{w_{t=1}}[w_{t=1}^1...w_{t=1}^N], where\ w=1/N$

$while\ s \leq t:$
$\quad \theta_s = \tilde{v}_{1:s-1}, \quad \lambda_s = \overline{u_{1:s-1}^2} \quad G=0$
$\quad for\ i \in [Ns:1]: w_{s|s-1}^i = w_{s-1}^i + \frac{\theta_s}{3}(w_{s-1}^{i-1} + w_{s-1}^i + w_s^{i+1})$
$\quad for\ i \in [1:Ns]: d_s^i = y_s - x_{s-i}, \quad \hat{w}_s^i = p(d_s^i | \lambda_s) w_{s|s-1}^i, \quad G = G + \hat{w}_s^i$
$\quad v_s = 0, \quad u_s = 0$
$\quad for\ i \in [1:Ns]: w_s^i = \hat{w}_s^i / G, \quad v_s = v_s + |w_s^i - w_{s-1}^i|, \quad u_s = u_s + w_{s-1}^i d_s^i$
$\quad s = s+1$

The implementation of the algorithm as described is then no more complex than running 3 Ns length loops. Note the necessity for the first for loop in the code to be reversed running from Ns to 1, this is due to the boundary condition that where, **w**$^{Ns+1}$ = 0, so without reversing the order of the loop the probability of the **w**$^{i+1}$ would always be 0.

A second implementation consideration is that over some lags will return infinitesimal probabilities – in most programming languages this will either cause an error as the floating point numbers overflow the memory limit of the language, or in some cases this will results in the probability being set to 0. A simple fix for this issue is to set a very small lower limit for the probability of each lag. Not doing this results in sub-optimal forecasts as the lag structure becomes increasingly path dependent as more lags are set to 0 over time.

**4. Experiments on Simulated Data**

In this section we are going to show the results of testing the SDM algorithm on a number of series constructed based on the assumptions of the Granger Causality model, as specified in (1), with the type of causality structures described by Cheung et al (2004) and Soros (2008). The purpose of this simulations is not to define exact mathematical models representative of these theoretical insights, rather to provide a number of cases that could arise in empirical work aimed at evidencing this area of financial theory.

In section 4.1. we will discuss the construction of these examples and present the equations used to generate the test series. We will then present results in section 4.2. of the root mean squared error forecasts we achieve using the SDM procedure to predict the values of the simulated price series, we show forecasts for a range of different noise levels.

4.1. *Construction of Simulated Series*

In the work of Cheung et al (2004) and Soros (2008) the notion of complex causality structures is only discussed in qualitative terms, this leaves no clear guidance as to what the best mathematical model to describe such series would be. To make sure we are covering a large range of possibilities we are going to test the SDM algorithm on 5 different series constructions. For each we are going to make the same assumptions as the Granger causality model described in (1), that the lagging series, in each of our example cases this will be the information series, is an autoregressive series constructed using;

$x_t = 0.9 x_{t-1} + N(0,1)$ (25)

where the autoregressive term is 0.9, so that the series exhibits significant but not infinite memory. This follows the testing framework used by Sornette and Zhou (2005).

As we don't know the functional form we would expect time-varying lags to take, we are going to use two simple functions. The first is a step function – so that the lag shifts between fixed regimes we specify in advance. The second is a random walk – so that the lag structure moves about arbitrarily within a bounded interval over time.

We are only going to focus on uni-directional forecasts, this greatly simplifies our results as otherwise we have to show two sets of statistics for each test, one for either series depending on which is leading in time. We note, however, that we would expect the same quality forecasts in bi-directional or positive-negative regime shifting cases as in the simple uni-directional case.

### 4.1.1. *Step Function Models*

We are going to construct two step function models. The first is a simple step function where there is a single lag at each t causally influencing the value of the price series. The second is a model where there are multiple lags influencing the price series. For the first model the price series $f_1(x)$ is constructed as;

$$f_1(x)_t = y_t = x_{t-\tau} + u_t \quad (26)$$

where the lag length τ varies as.

$$\tau_t = \begin{cases} 5 & if \quad 0 < t < 200 \\ 20 & if \quad 201 < t < 400 \\ 10 & if \quad 401 < t < 600 \end{cases} \quad (27)$$

The error term u is a mean 0 Gaussian noise, so that $\sigma_u$ is equal to the root mean square error (RMSE) we would expect from fitting a model of the x, y relationship if we knew the complete lag structure in advance.

For the second model, $f_2(x)$, we will also include the lagged values local to τ in the series construction i.e.;

$$f_2(x)_t = y_t = \frac{1}{7} \sum_{i=-3}^{i=3} x_{t-\tau+i} + u_t \quad (28)$$

So the price series is dependent on the 3 values of x immediately before and after τ in time.

### 4.1.2. *Random Walk Models*

We are also going to consider two models where the lag structure varies due to a random walk model. Here, we construct the series similarly to the step-function models but replace the lag-length function (27) with a trinomial random walk of the form;

$$\tau_t = \tau_{t-1} + z \quad \text{where} \quad f_z(z) = \begin{cases} 1/2, & z \in [-1, 0] & if \quad \tau \geq 25 \\ 1/3, & z \in [-1, 0, 1] & if \quad 5 < \tau < 25 \\ 1/2, & z \in [0, 1] & if \quad \tau \leq 5 \end{cases}, \quad (29)$$

indicating that the random walk is bounded so that the lag-length can only vary between 5 and 25 time-periods. The series $f_3(x)$ is then constructed similarly to (26) i.e. a random walk model where only 1 lag is causally influencing the price series. A second series $f_4(x)$ is constructed where the averaging function (28) is also applied to the series.

### 4.1.3. *Fixed-Integer-Lag Model*

A final model we consider is a fixed-integer-lag model where;

$$f_5(x)_t = y_t = x_{t-5} + u_t$$

we include this case as a baseline to show how well the algorithm forecasts in cases where there is no obfuscating lag variation.

### 4.1.4. Descriptive Statistics and Model Fitting

For each of the models we use the SDM algorithm exactly as specified in the pseudo-code example given in section 3.3. where we set Ns to 30 time-periods. To generate the forecast we take the probability weightings given to each of the lagging values, i.e. each of the values of $w_{t-1}$ and multiplying them by the corresponding values of **x** using.

$$\hat{y}_t = w_{t-1} \cdot x_{t-Ns:t-1} \quad (30)$$

We then calculate the RMSE for this forecast against the observed values of y as;

$$RMSE(\hat{y}, y) = \sqrt{\frac{1}{t} \sum_{i=1}^{i=t} (\hat{y}_i - y_i)^2} \quad (31)$$

Since we know that the expected error between the series is equal to $\sigma_u$ we also calculate a statistic for the deviation from expected forecast error as.

$$FE(\hat{y}, y) = RMSE(\hat{y}, y) - \sigma_u \quad (32)$$

### 4.2. Experimental Results

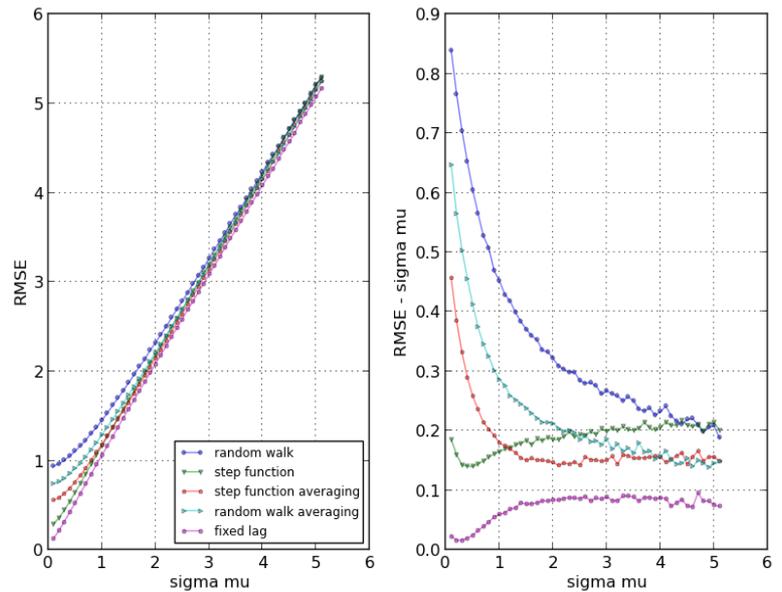

Figure 2 presents the results, each point on either plot represents the average over 500 trials for each model and noise level, $\sigma_\mu$. The left hand plot shows the RMSE in it's raw form calculated using (31). The right hand plot shows the FE statistic calculated using (32).

We can see from the left hand plot that the RMSE statistic tracks the value of $\sigma_\mu$ closely in all cases. For series where there is significant temporal variation, the random walk model $f_3(x)$ for example, we can see a period where the noise level is low $\sigma_\mu < 1$, there is a significant difference between the SDM model and the perfect fit. This is explicable by the fact that for these models, the expected fit does not take into account the extra noise from the temporal variation in the model structure.

*Figure 2: Experimental Results; the left hand plot shows the RMSE for the SDM model fit. The key indicates each of the different model types so $f_1(x)$ = step function, $f_2(x)$ = step function averaging, $f_3(x)$ = random walk, $f_4(x)$ = random walk averaging and $f_5(x)$ = fixed lag. The right hand plot shows the FE statistic for each model.*

The right hand plot also supports this theory, we see that for the fixed-integer-lag model SDM generated model fits very close to the maximum achievable < 0.1 standard deviations difference for all noise levels. As the temporal variance of the models increases, we see the initial (low noise) fits generated by the SDM algorithm getting increasingly poorer - although the model fits are clearly still very good.

## 5. Conclusions

The work of Cheung et al (2004) and Soros (2008) highlights a key issue with the way most financial forecasting research deals with time-sequencing. Participants in financial markets do not see the relationship between information and price as temporally fixed, but describe a situation where fluctuations in the timing of the information-price relationship are key drivers of the variation seen in asset prices. In the situation the authors describe understanding *when* information is likely to effect price is key to successful forecasting.

In this paper we have introduced a methodology that can capture the type of fluctuation the authors describe. We have shown that, under the same assumptions as the standard Granger causality approach to time-sequencing, the SDM algorithm produces the optimal Bayesian estimate of the forecasting distribution over the values of the lagging series. Importantly, the SDM algorithm we present will provide optimal Bayesian forecasts of the leading series in situations where the lag structure is not time-varying, but also in situations where it is.

We note that given the construction of the system equations we provide in this paper it would be possible to substitute the Gaussian measurement model we use for a range of other more complex models of bi-variate relationships present in the literature. The result is an estimation framework for time-varying lags, which is optimal under the conditions we describe, but also highly extensible to other cases.